# The Contagion Effects of Repeated Activation in Social Networks

Pablo Piedrahita[1,*], Javier Borge-Holthoefer[2], Yamir Moreno[1], and Sandra González-Bailón[3,*]

Abstract: Demonstrations, protests, riots, and shifts in public opinion respond to the coordinating potential of communication networks. Digital technologies have turned interpersonal networks into massive, pervasive structures that constantly pulsate with information. Here, we propose a model that aims to analyze the contagion dynamics that emerge in networks when repeated activation is allowed, that is, when actors can engage recurrently in a collective effort. We analyze how the structure of communication networks impacts on the ability to coordinate actors, and we identify the conditions under which large-scale coordination is more likely to emerge.

[1] University of Zaragoza
[2] Open University of Catalonia
[3] University of Pennsylvania

* Corresponding authors: ppiedrahita@gmail.com, sgonzalezbailon@asc.upenn.edu

Recent years have seen the emergence of massive events coordinated through large, decentralized networks. These include political protests and mobilizations like the Occupy movement of 2011 (Conover et al. 2013), the Gezi Park demonstrations of 2013 (Barberá et al. 2015), or the growth of the #BlackLivesMatter campaign during the 2014 protests in Ferguson (Freelon, McIlwain and Clark 2016). These collective events offer examples of the coordinating potential of communication networks – which, increasingly, emerge and take shape mediated by online technologies. This paper pays attention to the coordination dynamics that allow a small movement, a new campaign, or an unknown hashtag to rise to prominence. We present a formal model that allows us to answer the following question: How do coordination dynamics unfold to make individual actions (e.g. using an emerging hashtag, endorsing a mobilization) converge over time? Our model aims to disentangle the mechanisms that drive the emergence of decentralized, large-scale coordination. The goal is to identify the conditions under which coordination is more likely to arise from networks that are constantly pulsating with information.

Threshold models have become the standard for how we think about interdependence and the collective effects of social influence (Granovetter 1978; Granovetter and Soong 1983; Schelling 1978). As originally formulated, the activation of individual thresholds responds to global information, that is, to what everybody else in a collective is doing: the group of reference is assumed to be the same for all actors. In later developments of the basic model, networks were introduced to add local variance to social influence: the group of reference is now determined by connectivity in the network and it changes from actor to actor (Valente 1996; Watts 2002). These different variations of the threshold model, however, share two important elements: first, that activation is modelled as a step function that goes from 0 to 1 when thresholds are reached; and second, that thresholds can only be reached once, that is, activation is assumed to be a one-off event. Our model aims to relax these assumptions and allow actors to repeatedly activate as a function of the dynamics unfolding in the rest of the network. We argue that this modification aligns our model of contagion more closely with what is observed in many empirical networks – in particular, with the communication dynamics observed in online networks and the temporal autocorrelation that results from those dynamics.

Online campaigns are an important manifestation of this type of repeated activation. The Black Lives Matter movement, for instance, gained traction when the hashtag was first adopted in social media in 2013, which fueled what has been labeled as "an Internet-driven civil rights movement" (Eligon 2015, see also Day 2015). There is agreement that the movement consolidated with its first peaceful demonstrations in Ferguson in 2014 (Bosman and Fitzsimmons 2014). This "move from hashtag to the streets", claims one historian, took the movement "to a new phase" where it emerged as one of the most visible and coherent, "a model for how black liberations groups in the twenty-first century can organize an effective freedom rights campaign" (Ruffin 2015). Online networks were central to the coordination efforts of this campaign.

In the context of this empirical example, activation involves repeatedly using a specific hashtag to build momentum up to the point when large-scale coordination is achieved – and the hashtag receives global recognition. The goal of our model is to abstract this element and build an analytical framework around it to answer three interrelated questions: How does the structure of interdependence, the variance in individual propensities to activate, and the strength of social influence affect contagion and coordination? As with many analytical models, ours is a simplification of what is essentially a very complex reality. But it offers, we think, important insights into the counter-intuitive effects that networks have in allowing coordination to emerge.

The rest of the paper proceeds as follows. First, we consider prior work analyzing coordination in networks, and the analytical choices made when modelling social influence. We introduce our model as a continuation of threshold models, well suited to analyze dynamics of adoption (e.g. joining a political movement) but not well equipped to analyze the dynamics of coordination that emerge amongst actors that are already part of a movement. We then describe our model in detail, highlighting the main differences compared to previous approaches and unpacking our assumed mechanisms. In sections five and six we present our findings, which we organize around two main questions: How do changes in network topology affect the emergence of coordination under different assumptions of social influence? And how does individual heterogeneity impact coordination dynamics? We close the paper with a discussion of our findings, especially as they relate to previous research on contagion in networks.

## 1. Coordination as a Two-Step Selection Process

We can think of coordination dynamics as a two-step selection process: in the first stage, actors decide if they want to join a movement; in the second stage, they coordinate their actions with those who also opted in. Most threshold models refer to the first stage, and they focus on the cascading effects of one-off activations – the decision, that is, to join a collective action effort. Diversity in the motivation to become involved is modelled as a distribution of thresholds; what prior research shows is that the shape of this distribution is one of the key elements that explain the cascading effects of individual activations (Watts and Dodds 2010). Contagion dynamics, however, also depend on the structure of ties and how that structure encourages or hinders spreading dynamics. Networks shape coordination dynamics by creating different centrality distributions, which allow specific individuals to be more or less influential (Freeman, 1979); and by opening more or less structural holes (Burt 1992; also Girvan and Newman 2002), which constrain opportunities for chain reactions to the extent that they delimit the routes that cascades can follow (Watts 2002). Networks also delimit the size and the composition of the groups of reference surrounding a given actor, and therefore the number of social signals each actor receives (Centola and Macy 2007; Valente 1996).

Most threshold models assume that activation happens only once and that, once activated, the change of state (from inactive to active) is permanent. This is the reason why threshold models are particularly adept at capturing the first stage of coordination dynamics, for instance, the decision to join a movement or start using a particular hashtag. The model we propose here, on the other hand, aims to capture dynamics of activation within adoption, that is, coordination amongst actors who already opted in.

We have theoretical and empirical reasons to allow repeated activation to be the driving force of contagion dynamics. The empirical reason is that most instances of diffusion do not involve a single activation but many activations building up momentum in time. Before a hashtag becomes a trending topic, a period of buzz is first required; prior to a protest day, calls announcing the mobilization are distributed in waves. Actors decide whether they want to engage in an online conversation or take part in a protest. This is what threshold models can capture. What threshold models are not devised to capture is the period of information exchange that follows the act of joining a collective effort. During this period, social influence trickles intermittently as a function of the context that actors inhabit – that is, as a function of activity in the local networks to which they are exposed; and this context is not stationary: it changes, sometimes drastically, over time. Our model aims to capture this temporal dimension.

We also have a theoretical reason to relax the assumption of single activation. The intermittent dripping of information that social networks facilitate often leads to bursts of activity (Vazquez et al. 2006), as when news suddenly become trending topics (Lehmann et al. 2012; Wu and Huberman 2007). Coordination dynamics underlie these bursts of activity: sudden peaks in communication require the adjustment of individual actions, that is, the alignment of many individual decisions so that everybody uses the same trending hashtag or talks about the same news at the same time.

These dynamics of coordination, and how they lead to collective outcomes like swift information cascades, trending topics, or viral hashtags, are overseen if activation is modelled as a permanent change of state – that is, if we only focus on the first stage of what is, in fact, a two-step process. Our model presumes that, once in the second stage, individual propensities to activate will be influenced by the network and the signals it transmits, which in turn results from how other actors are influenced and react to that influence over time. These dynamics aim to resemble more closely the dynamics observed in the context of large-scale mobilizations, where actors repeatedly engage in activities like spreading calls for action or increasing the salience of political hashtags (Barberá et al. 2015; Borge-Holthoefer et al. 2011; Budak and Watts 2015; Conover et al. 2013; Jackson and Foucault Welles 2015). Individual decisions to contribute to the flow of information, and the decisions of those connected to a focal actor, co-evolve over time; our analytical approach models that co-evolution explicitly.

As previous models, our model assumes that exposure to information is the driving force underlying contagion. What makes our model different from previous models is that failure to trigger a chain reaction depends not only on the distribution of thresholds or the impact of network structure on activation dynamics; it also depends on whether the network facilitates coordination. By focusing on coordination dynamics, our model is in a better position to explain why, more often than not, large-scale contagion fails to take off. If the network is not conducive to coordination (i.e. if the timing of individual activations do not align over time), contagion ends up trapped in local activity clusters and, therefore, fails to synchronize the actions of the majority.

## 2. Model and Mechanisms

Our model of contagion relaxes the assumption that actors can only transition from an inactive to an active state. We also allow the effects of each activation to vary over time to the extent that they coevolve with the contagion dynamics taking place in the rest of the network. These modelling choices make sense if we think about how online networks facilitate contagion dynamics: users are constantly exposed to signals that might shift their inclination to act – for instance, send messages directing attention to specific issues (e.g. #occupy, #Gezi, #Ferguson, etc). Only when a large enough number of users converge in their attention to these issues, their actions become globally visible – i.e. mass media starts paying attention. This type of coordination not only affects trending buzz; it actually has the potential to shape the public agenda in the same way than more traditional social movements would (Petersen-Smith 2015). The difference is that coordination happens spontaneously, from the bottom-up.

To bring these empirical intuitions into a tractable framework, we follow classic models of synchronized coordination (Mirollo and Strogatz 1990; Piedrahita et al. 2013). These models have been used extensively to study coordination dynamics in biological and physical settings (Strogatz 2003), but they have never been used, to the best of our knowledge, to illuminate dynamics relevant for the study of social mobilization, or to extend classic threshold models and their application to sociological questions. Like threshold models, our model assumes that the motivational structure of actors can be defined by a limit that, when reached, triggers activation; unlike threshold models, we split the motivation to activate into two components: a social component, which depends on what other actors are doing; and an individual component, which defines the intrinsic propensity of actors to activate regardless of what others are doing. We model this intrinsic component as a function that increases monotonically over time within the range [0,1] until the upper bound – which acts as the threshold or activation limit – is reached. Figure 1 illustrates the logic of this approach.

-- Figure 1 about here --

Our main assumption is that actors reach their activation zone at different speeds. The speed of activation is a function of two parameters: $\omega$, which determines how quickly the actor reaches the threshold zone (i.e. it defines the concavity of the curve that maps progression towards activation); and $\varepsilon$, or the strength of the signal received from other actors – which, in our case, is restricted to actors one step removed in the network. Every time a neighbor activates, they send a pulse that shifts the state of the focal actor closer towards the threshold zone; the parameter ε, in other words, is the building block we use to introduce social influence in the model. The lower panels in figure 1 illustrate how actor *i* advances towards activation, both as a function of her intrinsic propensity ω and as a response to the activation of the neighbors. When a node activates, as node *i* does in t2, she shifts the state of her neighbors with the ε signal and resets her state back to the beginning of her phase.

The mathematical expression of these intuitions follows this functional form:

$$x = f(t) = \frac{1}{\omega} \ln\left(1 + [e^{\omega} - 1]t\right) \tag{1}$$

The parameter $\omega$ determines the shape of this function. It is always ω > 0 to make the function concave down. We assume a monotonic increase because it is the most natural choice when modeling progression towards activation and it follows the same intuition as threshold models – only that instead of proposing a stepwise change, it models actors' propensity to activate as a continuous progression. As figure 2 shows, a larger ω produces a more pronounced shape, making the function rise very rapidly to then level off.

-- Figure 2 about here --

In addition to this individual component, the model also takes into account the activation of other actors in the network, in particular, those one step removed. If actors *i* and *j* are connected, *j*'s activation increases *i*'s propensity to activate by an amount ε or pushes *i* directly into activation, whichever is less. This rule of interdependence is expressed as:

$$x_i = \min(1, x_i + \varepsilon) \tag{2}$$

The parameter ε captures social influence. In our model, the activation signals sent by neighbors are more consequential if they are concurrent (as in panel t4 of figure 1) than if they are not (panel t3). In other words: our model assumes that exposure to multiple signals matters not just because it reinforces affirmation (a process that we capture with the sudden increases in the progression towards the threshold zone specified by equation 2); but also, and mostly, because it allows local activity to grow increasingly correlated over time. Introducing this temporal correlation is, we believe, a necessary ingredient to build realistic models of large-scale coordination, especially given the available evidence on the temporal dynamics and bursts of activity characteristic of human communication (Vazquez et al. 2006).

Our analytical choice acknowledges the important difference between having multiple friends participating in, say, #BlackLivesMatter discussions or encouraging #OccupyCentral actions at different, uncoordinated times than having them all converge to the same timing. Convergence in the timing of activations is more conducive to further activations, which, in turn, reinforces the feedback mechanism that makes an obscured issue suddenly jump to the spotlight of media attention. This is what happened in Ferguson during the first hours of the demonstrations. Journalists learned about the events through their Twitter feeds (where hashtags created a channel for relevant information to flow in a coordinated fashion), not from their own news organizations (Carr 2014). Large-scale coordination becomes visible only when the timings of individual activations become highly correlated; and this is an aspect that cannot be captured by models that disregard the effects of time on activation dynamics.

To sum up, the motivational structure of actors in our model is determined by the parameter $\omega$, which defines how quickly they reach the activation zone; and by the parameter $\varepsilon$, which determines the strength of social influence. In a world of isolated actors, $\varepsilon$ would equal 0; in a world where social influence overrides the rhythms of intrinsic activation, $\varepsilon$ would equal 1. Likewise, in a world of identical actors, $\omega$ would be distributed homogeneously; the more heterogeneous the distribution, the more unequal actors are in their propensity to activate. These two parameters open the basic experimental space of our model.

## 3. The Dynamics of Repeated Activation

The combination of values for the two parameters $\omega$ and $\varepsilon$ (when $\varepsilon > 0$) determines the speed at which coordination emerges – that is, how long it takes for all nodes to start pulsating, or activating, concurrently (e.g. many people simultaneously using a new hashtag). However, the underlying network determining the pathways for influence is also a crucial component of how we think about contagion dynamics. The core of our analyses aim, in fact, to determine the impact that different network structures have on those dynamics, holding $\omega$ and $\varepsilon$ constant.

To illustrate why networks matter in the context of our model, figure 3 summarizes contagion dynamics in toy networks with size N = 10, and values $\omega = 3$ and $\varepsilon = 0.008$. The initial state randomly allocates actors to different points in their progression towards activation, which means that they pulsate at different times, as matrices $t_{cycle} = 1$ show (a $t_{cycle}$ is complete when every actor in the network has activated at least once). As time progresses, however, the impact of social influence (via $\varepsilon$) starts aligning nodes to the same timing. This is particularly clear in the case of the directed cycle. The undirected version of the cycle requires more time for actors to coordinate their activations; in fact, there is still an actor that activates with its own timing at $t_{cycle} = 70$. The tree structure, also undirected, is the least conducive to coordination: the presence of hubs, and their greater influence over the peripheral nodes that are only connected through them, hampers the spontaneous emergence of coordination.

-- Figure 3 about here –

## 4. Social Context as Communication Networks

The topology on which interactions take place is, therefore, a crucial element in the dynamics we want to model. We run experiments on four network topologies, summarized in figure 4. These networks determine how actors influence each other via the ε signal, and they capture different hypothetical scenarios where interactions might unfold empirically.

-- Figure 4 about here --

In the Erdős–Rényi network, for instance, ties connecting the actors are formed at random. Although we know that social networks are never formed at random, this topology could account for a scenario where actors are connected through their online search patterns, i.e. by looking at what others are posting on websites or blogs beyond social media platforms. This network also offers a standard benchmark with which to assess the performance of the other three topologies.

The regular network offers a way of mapping interdependence when it is highly structured by logistical or space constraints. During the 2014 Umbrella Revolution in Hong Kong, for instance, social media and other Internet-based modes of communication were censored by the Chinese government, so protesters used the Bluetooth technologies in their cell phones to create mesh networks and coordinate their actions while on the streets (Knibbs 2014; Parker 2014; Rutkin and Aron 2014). These networks do not rely on online servers (and are, therefore, more difficult to monitor by third parties); but they require physical proximity: they are only feasible when there is a large number of people concentrated in restricted spaces (like concert halls, stadiums or, as in this case, a few streets within the same city district). Regular networks offer an approximation to that sort of empirical scenario.

The small world and scale free networks are the topologies we use to approximate most observed networks. There is ample evidence that social networks exhibit the small world property (Watts 2003) and they also tend to have a very skewed degree distribution, especially those that emerge online (Barabási 2009). Twitter, for instance, has a long tail in the allocation of connections, with a minority of accounts being disproportionately better connected than the vast majority (Kwak et al. 2010). Similar properties have been found in other social media platforms like Facebook or the Chinese Sina Weibo (Backstrom et al. 2012; Ugander et al. 2011; Zhengbiao, Zhitang and Hao 2011). We reproduce these structural features in our experiments because social media networks have been shown to play an important part in the emergence of large-scale coordination, from agreeing on which hashtags to use to organizing massive demonstrations (Barberá et al. 2015; Conover et al. 2013; González-Bailón et al. 2011; Romero, Meeder and Kleinberg 2011; Steinert-Threlkeld et al. 2015). To recover the example introduced above, the growth of the #BlackLivesMatter movement relied heavily on the coordinating potential of social media.

## 5. The Effects of Network Topology

The main motivation guiding our analyses is to identify the conditions that need to be in place for large-scale coordination to emerge. Given that the time to full coordination depends on the specific combination of $\omega$ and $\varepsilon$, but also on the underlying network, we measure time in terms of $t_{cycles}$, which were illustrated in figure 3. This definition allows us to normalize time across conditions and directly compare coordination dynamics across networks and parametric settings. From an empirical point of view, every step in the evolution of our model (every $t_{cycle}$) can be interpreted as a different time window, e.g. hourly, daily, weekly, or monthly activity. Finding the appropriate temporal resolution to empirically analyze evolving dynamics in networks is not a trivial issue (Holme and Saramäki 2012; Moody 2002). Our model does not make any specific assumptions about the right resolution to aggregate observed activation data; the time it takes for a cycle to complete can correspond to different empirical windows – and, in fact, the appropriate width for that window is likely to change as periods of bursts in activity unfold in chronological time (Borge-Holthoefer et al. 2016).

At the end of every $t_{cycle}$, that is, once every node has activated at least once, we count the number of nodes that activated simultaneously – i.e. the size of the clusters in the matrices of figure 3. Our model allows large-scale coordination to arise when small local islands of coordinated nodes start merging together through the cascading effects of social influence, as captured by the parameter ε and as channeled by the network. Figure 5 shows what happens with the levels of coordination as the system evolves with a fixed $\varepsilon = 0.01$ on a small world graph with size $N = 10^4$. The curves track the fraction of actors that activate simultaneously. As expected, high ω values (which, in this example, is the same for all actors) lead to faster large-scale coordination. As ω decreases, the time to full coordination increases. At low values (i.e. $\omega = 6$), system-level coordination is unattainable: this is a condition under which only small clusters of coordinated nodes emerge.

-- Figure 5 about here --

The question we are interested in is: how do contagion dynamics differ when ω and ε are held constant but the underlying networks change? Figure 6 shows a first set of results to answer this question. Every dot in the heatmaps corresponds to a combination (ε, ω). On the left of the horizontal axis we have systems where social influence is very strong; as we move to the right, the impact of neighbor activations on the focal actor starts diminishing. At the bottom of the vertical axis, we have actors that progress slowly towards the activation zone; at the top, we have those that get very quickly into a tipping-point state. In this set of simulations, the propensity to activate (the ω value) is distributed homogeneously across all actors in the network; what changes is the structure of the underlying network.

The color scheme indicates the time it takes under each parametric combination to reach large-scale coordination, measured as $t_{cycles}$. We define large-scale coordination as having at least 75% of the nodes activating simultaneously. For each point, time is averaged over 100 realizations of the simulation, with different initial conditions. In this scheme, lighter colors indicate earlier coordination; as the colors get darker, coordination takes longer to emerge. Black signals that no coordination was possible within the limit of 200 $t_{cycles}$, when the simulations stopped.

-- Figure 6 about here --

These results suggest that all networks are capable of generating coordination in scenarios with strong to moderate social influence (0.5 < ε < 1), regardless of the actors' propensity to activate (regardless of the ω value). As ε starts getting smaller (i.e. as the strength of social influence diminishes), actors need to have steeper inclinations to reach the tipping point for coordination to emerge. A network where ties channel little impact takes more time, and requires more motivated actors, to generate the same level of coordination than a network with stronger ties. After some critical point, no amount of actor predisposition can overcome the lack of substantive social influence. This critical point, however, changes across networks: in the random, Erdős–Rényi network, large-scale coordination emerges for most social influence conditions when ω is high, even when the impact of each neighbor activation is really low. This is not the case for the regular, the small world, and the scale free networks, which are way more restrictive in their support to spontaneous coordination. The scale-free network is particularly limiting: it either allows coordination to emerge fast (white region) or it prevents it very abruptly (black region). The existence of hubs, so characteristic in the structure of these networks, explains why such an abrupt transition takes place: because hubs are so much better connected than the other nodes, they have a wide impact when they activate; but hubs, which are surrounded by many structural holes (Burt 1992), also restrict the pathways for contagion, and for the alignment of local dynamics.

Given that most social media networks are well represented by the scale free structure, our simulations suggest two possibilities: either online ties channel stronger influence than traditionally acknowledged (e.g., Gladwell 2010); or users are so ready to activate that coordination is possible even with weak social influence (but not too weak). This is indeed what seems to happen during the emergence of campaign hashtags. Social media users tend to be proactive in their behavior to facilitate coordination; in fact, the use of hashtags in Twitter emerged itself as a user-driven convention (see Parker 2011). This high predisposition, especially amongst those who opted into a movement or mobilization (as our model presumes), compensates for the hurdles imposed by the network to spreading dynamics.

## 6. The Effects of Actor Heterogeneity

The findings above are interesting because they cast light on the importance that network topology has to delimit the possibility space for large-scale coordination. However, it is a big simplification to assume that all actors have the same propensity to reach their activation zone. In a second set of simulations, we introduced heterogeneity in the distribution of the ω parameter, as illustrated in figure 7. We randomly drew $N = 10^4$ values from a normal distribution centered around $\omega = 50$ and a standard deviation in the interval $\sigma = [1,10]$, with 0.1 increases. A condition where actors differ slightly in their predispositions to act corresponds to scenarios where exogenous events instill a sense of urgency in the need to act, as it happened in Ferguson. A condition where actors are very heterogeneous, on the other hand, corresponds to situations where the level of commitment to a cause varies amongst those willing to participate. For instance, in the Hong Kong protests students triggered a movement that soon escalated to involve a larger group of participants, partly thanks to the aid of social media (Parker 2014). Students had the ability to camp on the streets and the time to generate the messages, photos, and videos that others (including mainstream media) later picked up. Other demographic groups (parents, middle class professionals) might have wanted to join the protests but they were unable to do so with equal intensity because of their job schedules or family constraints. Sociological factors like these could be a source of heterogeneity in the ability to activate for actors that are, otherwise, equally interested in a political cause.

-- Figure 7 about here --

The results of this second set of experiments are shown in figure 8. In general, the simulations reveal that heterogeneity reduces opportunities for large-scale coordination across all networks. This supports the intuition that, for a cause to grow large, actors need to share predispositions, that is, they need to be as similar as possible in their willingness to act. The scale-free network is, again, the most restrictive structure – but as long as ties channel some influence, coordination arises fast, which is important for time-sensitive mobilizations (for instance, during the first hours of the 2013 Gezi Park protests, when mainstream media were censorig news of the events on the ground, Barberá et al. 2015). Given that large-scale coordination emerges repeatedly (and swiftly) in social media sites, our simulation results provide further evidence that online ties weave relevant interdependence, that is, they act as a significant source of social influence. This is consistent with experimental evidence on the mobilizing potential of online networks (Bond et al. 2012), which shows that exposure to information through social media has a positive and significant impact on political behavior. This positive impact is what we capture with the ε parameter. Our results show that as long as the impact of social influence is not too low, it can drive the network towards coordination even when actor heterogeneity is high.

-- Figure 8 about here –

## 7. Discussion

Overall, our results show that network topology has counter-intuitive effects on coordination when repeated activation is allowed. Homogeneous networks, that is, networks where the degree distribution is not significantly skewed, are more conducive to coordination: the parametric combinations (ω, ε) leading to coordination are wider for more egalitarian networks, following this order: Erdős–Rényi > small world > regular networks. This ranking applies to conditions where ω is fixed but also where it is distributed randomly. Networks characterized by a skewed degree distribution (that is, by the presence of a small group of nodes exerting more influence over other nodes) are clearly less favorable to coordination: they require stronger influence and actors that are more similar in their propensity to activate.

We refer to this as the "scale-free paradox": on the one hand, scale-free networks clearly create worse conditions for contagion dynamics to spread under repeated activation; on the other hand, an increasing body of observational evidence shows that these networks are also very good at helping coordinate the actions of many (Barberá et al. 2015; Conover et al. 2013; González-Bailón et al. 2011; Romero, Meeder and Kleinberg 2011; Steinert-Threlkeld et al. 2015). This empirical evidence suggests that heterogeneous networks are indeed behind many observed episodes of mass mobilization, regardless of the topological restrictions uncovered by our simulation results.

Related to this, our findings also suggest that online networks must channel enough social influence to allow individual actions to align over time. Our results show that there is a critical $\varepsilon$ for all topologies, that is, for a given network and $\omega$ there is always a value for the social influence parameter below which actors do not achieve coordination. This critical value changes across topologies, and it is particularly stringent for the scale-free networks. Since most online networks are skewed in their degree distribution, we contend that those networks must channel moderate to strong social influence – otherwise, it is unlikely that large-scale coordination would emerge so often though online channels.

Time-varying dynamics in networks have so far been largely disregarded by analytical approaches to collective action – and yet these dynamics are crucial, as our model suggests, to understand the feedback mechanisms that activate cascading reactions and the consolidation of a critical mass. Prior research has shown that attaining this critical mass depends on the network topology, in particular the density and the centralization of ties (Marwell and Prahl 1988). That work suggests that centralization always has a positive effect on collective action because it increases the probability that involved actors will be tied to a large number of contributors, allowing for more efficient coordination. Our model suggests that highly centralized networks (in the form of scale-free structures) can indeed be very efficient in coordinating efforts but only when certain conditions are met. The strength of social influence, and the distribution of propensities to activate, need both to be conducive to the critical mass. Compared to other network topologies, however, centralized structures perform significantly worse, all else equal.

By allowing activation to re-occur, we shift attention from the diffusion of activations (the focus of traditional threshold models) to their coordination (which happens during the second stage of activity within adoption). What we find is that for a range of parametric combinations ($\omega$, $\varepsilon$), the four network topologies we analyze are equally successful at generating coordination. What makes them differ is the impact that social influence has on collective dynamics. As networks grow more heterogeneous in their connectivity, and as they open more structural holes, the space for the emergence of large-scale coordination diminishes. This difference across networks results from how the underlying structure of communication activates feedback mechanisms of reinforcement that align, with more or less success, individual decisions to activate.

There are two aspects of our modeling approach that deserve future consideration: the distribution of $\varepsilon$ (which we keep constant across ties) and the way in which $\omega$ values are distributed (randomly, when heterogeneous).There are a number of reasons why these two choices could be modified. We know that in social networks ties vary in their strength: the actions of relatives, friends and acquaintances, for instance, do not have the same effect on an actor's behavior. Our model assumes that all ties channel the same amount of influence. Although some ties activate more often than others (and are *de facto* more influential), their impact on activation responds to changing local events in the network, not to an attribute of the tie itself. Future work should consider coordination dynamics under different distributional assumptions of ε. Likewise, future research should analyze scenarios where the propensity to activate (ω) is not distributed randomly but as a function of the network topology itself. For instance, there is ample empirical evidence to suggest that the values of ω might be more similar within clusters in a network – if we assume that this is another dimension on which homophily operates (McPherson, Smith-Lovin and Cook 2001). Students, for example, are more likely to share the same predispositions and be better connected to each other compared to other demographic groups. Critical mass dynamics and the timing of coordination are likely to differ if we constrain the distribution of ω to the position of nodes in the network.

     Another important question for future research is how much the results would vary if actors were equipped with memory, that is, if they did not reset their progression towards activation to 0 at the end of every $t_{cycle}$. Equipping actors with memory would open the door to more explicit theorization on the impact that mechanisms like social learning have on activation dynamics. As it currently stands, our model is aseptic about the specific mechanisms that give shape to the function expressed in equation (1). Our main explanatory variables are the networks assumed to underlie coordination dynamics; we treat ω as a black box that determines the timing of individual activations. When we allow ω to differ from actor to actor, the assumed heterogeneity can relate to different empirical possibilities: more or less interest in a political cause, more or less time to devote to the cause, etc. In any case, adding memory to how our actors behave would require a solid empirical justification of how memory operates in the context of coordination through decentralized networks.

Finally, another important question that we do not consider directly relates to finding a temporal scale that is the most appropriate to empirically analyze coordination dynamics. As with most analytical models, ours is developed on a level of abstraction that allows generalizing across possible scenarios but does not give precise guidelines as to how to aggregate empirical data. Digital technologies are providing richer sources of data that could help test empirically models like ours (Golder and Macy 2014; Lazer et al. 2009; Watts 2007). Our model, in particular, requires a systematic approach to the analysis of time-evolving networks and time-dependent activations (Holme and Saramäki 2012; Moody 2002). In data tracking social media activity, the temporal scale can be expressed in terms of days, hours, or minutes – and the most informative temporal scale might not even remain constant during the observation window (Borge-Holthoefer et al. 2016). Bringing closer the results of simulation models with the patterns observed in empirical data requires solving first the temporal resolution problem. More research is necessary in this area.

## 8. Conclusion

The model presented here casts light on how contagion dynamics emerge when actors are allowed to activate repeatedly and contribute intermittently to activity around a collective cause. Theories of a critical mass and threshold models emphasize the importance of interdependence, and highlight that collective action is not about obtaining unanimous participation but about mobilizing enough people to make the effort self-sustaining. Our model contributes to this broad line of research by focusing on the second stage of coordination within adoption, that is, on the exchange of information among actors who are already part of a political cause. We emphasize the importance of temporal correlations in network activity, so far largely disregarded in previous modelling efforts but characteristic of many recent examples of observed large-scale coordination. Our model shows that many contagion conditions are not conducive to coordination. In particular, networks that are more homogenous in their degree distribution facilitate coordination under a wider range of actor predisposition and social influence conditions; as inequality in the degree distribution increases, however, so does the time required to achieve coordination – time that, from an empirical point of view, might not always be available. Our model also shows that when social influence has a moderate to strong impact, large-scale coordination emerges regardless of the underlying structure of communication, and regardless of actor's predisposition to act. To the extent that digital technologies are inserting networks in every aspect of social life, our results suggest that we should expect to see more instances of large-scale coordination cascading from the bottom-up.

References

Backstrom, Lars, Paolo Boldi, Marco Rosa, Johan Ugander, and Sebastiano Vigna. 2012. "Four degrees of separation." Pp. 33-42 in *Proceedings of the 3rd Annual ACM Web Science Conference*. Evanston, Illinois: ACM.

Barabási, Albert-László. 2009. "Scale-Free Networks: A Decade and Beyond." *Science* 325(5939):412-13.

Barberá, Pablo, Ning Wang, Richard Bonneau, John Jost, Jonathan Nagler, Joshua Tucker, and Sandra González-Bailón. 2015. "The Critical Periphery in the Growth of Social Protests." *PloS ONE* 10(11).

Bond, Robert M., Christopher J. Fariss, Jason J. Jones, Adam D. I. Kramer, Cameron A. Marlow, Jaime E. Settle, and James H. Fowler. 2012. "A 61-Million-Person Experiment in Social Influence and Political Mobilization." *Nature* 489:295-98.

Borge-Holthoefer, Javier, Nicola Perra, Bruno Gonçalves, Sandra González-Bailón, Alex Arenas, Yamir Moreno, and Alessandro Vespignani. 2016. "The dynamics of information-driven coordination phenomena: a transfer entropy analysis." *Science Advances* 2(4).

Borge-Holthoefer, Javier, Alejandro Rivero, Iñigo García, Elisa Cauhé, Alfredo Ferrer, Darío Ferrer, David Francos, David Iñiguez, María Pilar Pérez, Gonzalo Ruiz, Francisco Sanz, Fermín Serrano, Cristina Viñas, Alfonso Tarancón, and Yamir Moreno. 2011. "Structural and Dynamical Patterns on Online Social Networks: the Spanish May 15th Movement as a Case Study doi:10.1371/journal.pone.0023883." *PloS ONE* 6(8):6e23883.

Bosman, Julie, and Emma G. Fitzsimmons. 2014. "Grief and Protests Follow Shooting of a Teenager." in *The New York Times*.

Budak, Ceren, and Duncan J. Watts. 2015. "Dissecting the Spirit of Gezi: Influence vs. Selection in the Occupy Gezi Movement." *Sociological Science* 2(7).

Burt, Ronald S. 1992. *Structural Holes. The Social Structure of Competition*. Cambridge, MA: Harvard University Press.

Carr, David. 2014. "View of #Ferguson Thrust Michael Brown Shooting to National Attention." in *The New York Times*.

CONTAGION EFFECTS OF REPEATED ACTIVATION    25

Centola, Damon, and Michael W. Macy. 2007. "Complex Contagions and the Weakness of Long Ties." *American Journal of Sociology* 113(3):702-34.

Conover, Michael D., Emilio Ferrara, Filippo Menczer, and Alessandro Flammini. 2013. "The Digital Evolution of Occupy Wall Street." *PloS ONE* 8(5):e64679.

Day, Elizabeth. 2015. "#BlackLivesMatter: the birth of a new civil rights movement." in *The Guardian*.

Eligon, John. 2015. "One Slogan, Many Methods: Black Lives Matter Enters Politics." in *The New York Times*.

Freelon, Deen, Charlton McIlwain, and Meredith Clark. 2016. "Quantifying the power and consequences of social media protest." *New Media & Society*:1461444816676646.

Freeman, Linton C. 1979. "Centrality in Social Networks: Conceptual clarification." *Social Networks* 2(3):215-39.

Girvan, M., and M.E.J. Newman. 2002. "Community structure in social and biological networks." *Proceedings of the National Academy of Sciences* 99(12):7821-26.

Gladwell, Malcolm. 2010. "Small Change. Why the revolution will not be tweeted." *The New Yorker*.

Golder, Scott A., and Michael W. Macy. 2014. "Digital Footprints: Opportunities and Challenges for Online Social Research." *Annual Review of Sociology* 40(1):129-52.

González-Bailón, Sandra, Javier Borge-Holthoefer, Alejandro Rivero, and Yamir Moreno. 2011. "The Dynamics of Protest Recruitment through an Online Network." *Scientific Reports* 1:197.

Granovetter, Mark. 1978. "Threshold Models of Collective Behavior." *American Journal of Sociology* 83(6):1420-43.

Granovetter, Mark, and Roland Soong. 1983. "Threshold Models of Diffusion and Collective Behaviour." *Journal of Mathematical Sociology* 9:165-79.

Holme, Petter, and Jari Saramäki. 2012. "Temporal networks." *Physics Reports* 519(3):97-125.

Jackson, Sarah J., and Brooke Foucault Welles. 2015. "Hijacking #myNYPD: Social Media Dissent and Networked Counterpublics." *Journal of Communication* 65(6):932-52.

Knibbs, Kate. 2014. "Protesters are using FireChat's Mesh Networks to Organize in Hong Kong." in *Gizmodo*.

Kwak, Haewoon, Changhyun Lee, Hosung Park, and Sue Moon. 2010. "What is Twitter, a Social Network or a News Media?" in *Proceedings of the 19th International World Wide Web Conference (WWW 2010)*.

Lazer, David, Alex Pentland, Lada Adamic, Sinan Aral, Albert-László Barabási, Devon Brewer, Nicholas A. Christakis, Noshir S. Contractor, James H. Fowler, Myron Gutmann, Tony Jebara, Gary King, Michael W. Macy, Deb Roy, and Marshall Van Alstyne. 2009. "Computational Social Science." *Science* 323:721-23.

Lehmann, Janette, Bruno Goncalves, Jose J. Ramasco, and Ciro Cattuto. 2012. "Dynamical classes of collective attention in twitter." Pp. 251-60 in *Proceedings of the 21st international conference on World Wide Web*. Lyon, France: ACM.

Marwell, Gerald, and Ralph Prahl. 1988. "Social networks and collective action. A theory of critical mass III." *American Journal of Sociology* 94:502-34.

McPherson, M., L. Smith-Lovin, and J. Cook. 2001. "Birds of a feather: Homophily in Social Networks." *Annual Review of Sociology* 27:415-44.

Mirollo, R., and S. Strogatz. 1990. "Synchronization of Pulse-Coupled Biological Oscillators." *SIAM Journal on Applied Mathematics* 50(6):1645-62.

Moody, James. 2002. "The Importance of Relationship Timing for Diffusion." *Social Forces* 81(1):25-56.

Newman, M.E.J. 2010. *Networks. An Introduction*. Oxford: Oxford University Press.

Parker, Ashley. 2011. "Twitter's Secret Handshake." in *The New York Times*.

Parker, Emily. 2014. "Social Media and the Hong Kong Protests." *The New Yorker*.

Petersen-Smith, Khury. 2015. "Black Lives Matter: a New Movement Takes Shape." *International Socialist Review* (96).

Piedrahita, Pablo, Javier Borge-Holthoefer, Yamir Moreno, and Alex Arenas. 2013. "Modeling self-sustained activity cascades in socio-technical networks." *EPL (Europhysics Letters)* 104(4):48004.

Romero, Daniel M., Brendan Meeder, and Jon Kleinberg. 2011. "Differences in the Mechanics of Information Diffusion Across Topics: Idioms, Political Hashtags, and Complex Contagion on Twitter." in *International World Wide Web Conference*. Hyderabad, India.

Ruffin, Herbert. 2015. "Black Lives Matter: The Growth of a New Social Justice Movement." in *BlackPast.org*.

Rutkin, Aviva, and Jacob Aron. 2014. "Hong Kong Protesters Use a Mesh Network to Organize." in *NewScientist*.

Schelling, Thomas C. 1978. *Micromotives and Macrobehavior*. London: Norton.

Steinert-Threlkeld, Zachary C, Delia Mocanu, Alessandro Vespignani, and James Fowler. 2015. "Online social networks and offline protest." *EPJ Data Science* 4(1):19.

Strogatz, S. H. 2003. *SYNC: the emerging science of spontaneous order*. New York, NY: Theia.

Ugander, Johan, Brian Karrer, Lars Backstrom, and Cameron A. Marlow. 2011. "The Anatomy of the Facebook Social Graph." *arXiv* 1111.4503v1.

Valente, Thomas W. 1996. "Social network thresholds in the diffusion of innovations." *Social Networks* 18:69-89.

Vazquez, Alexei, Joao Gama Oliveira, Zoltan Dezso, Kwang-Il Goh, Imre Kondor, and Albert László Barabási. 2006. "Modeling bursts and heavy tails in human dynamics." *Physical Review E 73* 036127:1-19.

Watts, Duncan J. 2002. "A Simple Model of Global Cascades on Random Networks." *PNAS* 99:5766-71.

—. 2003. *Six Degrees. The Science of a Connected Age*. London: William Heinemann.

—. 2007. "A twenty-first century science." *Nature* 445:489.

Watts, Duncan J., and Peter S. Dodds. 2010. "Threshold Models of Social Influence." in *Handbook of Analytical Sociology*, edited by Peter Bearman and Peter Hedström. Oxford: Oxford University Press.

Watts, Duncan J., and Steven H. Strogatz. 1998. "Collective dynamics of 'small world' networks." *Nature* 393(4):440-42.

Wu, Fang, and Bernardo A. Huberman. 2007. "Novelty and collective attention." *Proceedings of the National Academy of Sciences* 104(45):17599-601.

Zhengbiao, Guo, Li Zhitang, and Tu Hao. 2011. "Sina Microblog: An Information-Driven Online Social Network." Pp. 160-67 in *Cyberworlds (CW), 2011 International Conference on*.

Figure 1. Schematic Representation of the Social Influence Model with Recurrent Activation

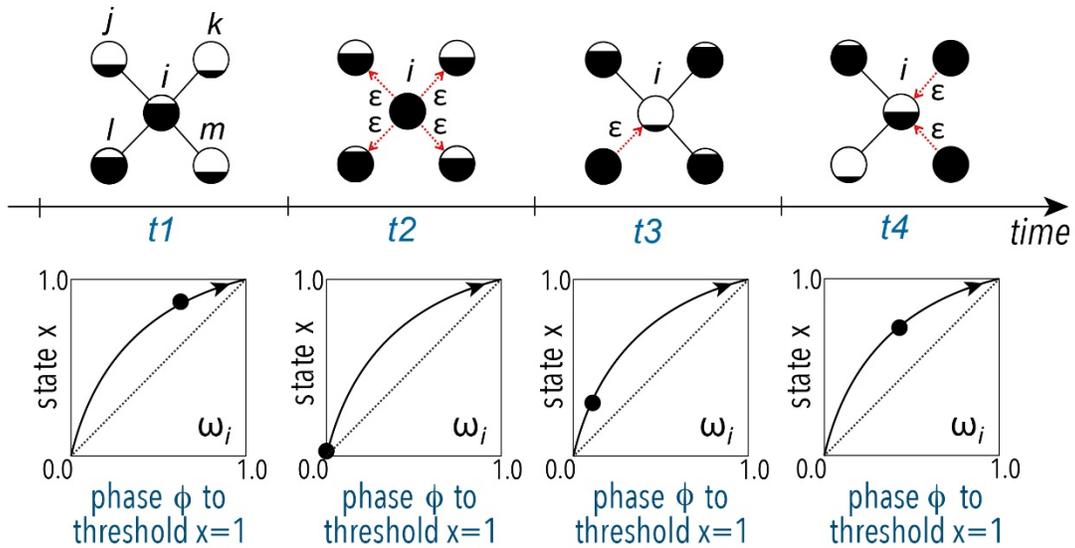

Note: The model, adapted from (Mirollo and Strogatz 1990), assumes that actors (i.e. the nodes in the network) reach their activation threshold at different speeds. The speed of activation is a function of two parameters: ω, which determines how quickly the actor reaches the threshold zone (i.e. it defines the concavity of the curve that maps progression towards activation); and ε, or the strength of the signal received from the neighbors in the network when they activate, a pulse that shifts the state of the focal actor closer towards the threshold (the timing of which varies over time). The lower panels in the figure illustrate how actor $i$ advances towards activation. When a node activates, as node $i$ does in t2, she shifts the state of her neighbors with the ε signal and resets her state back to the beginning of her phase.

Figure 2. The Impact of the Parameter ω on the Activation Buildup

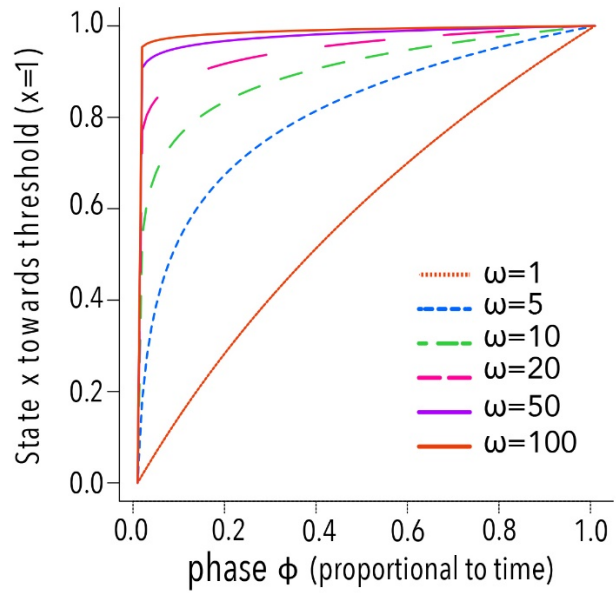

Note: When the parameter ω is 0, the progression of actors towards their activation threshold ($x = 1$) grows linearly with time; as the parameter ω increases, actors reach their activation zone faster, i.e. a signal received from their neighbors will tip their sate over the threshold, which means they will send a signal as well (thus helping other actors also get closer to their activation zones).

Figure 3. The Impact of Network Topology on Coordination Dynamics

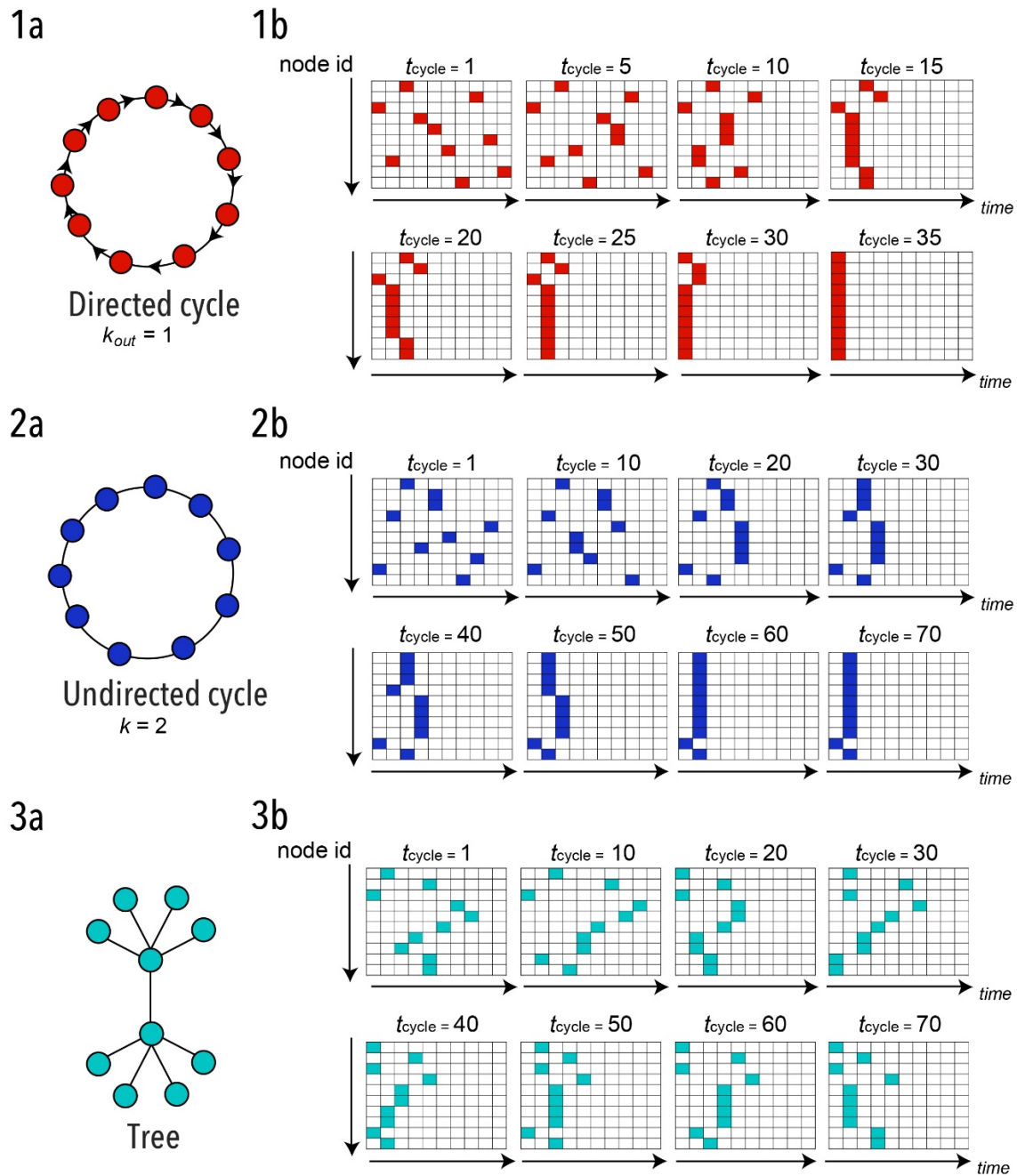

Note: This figure summarizes contagion dynamics in toy networks with size $N = 10$, and values $\omega = 3$ and $\varepsilon = 0.008$. The initial state randomly allocates actors to different points in their progression towards activation. As time passes, the impact of social influence (via $\varepsilon$) starts aligning nodes to the same timing (a $t_{cycle}$ is complete when every actor in the network has activated at least once). The tree structure is the least conducive to coordination.

Figure 4. Network Topologies Used in the Simulation Experiments

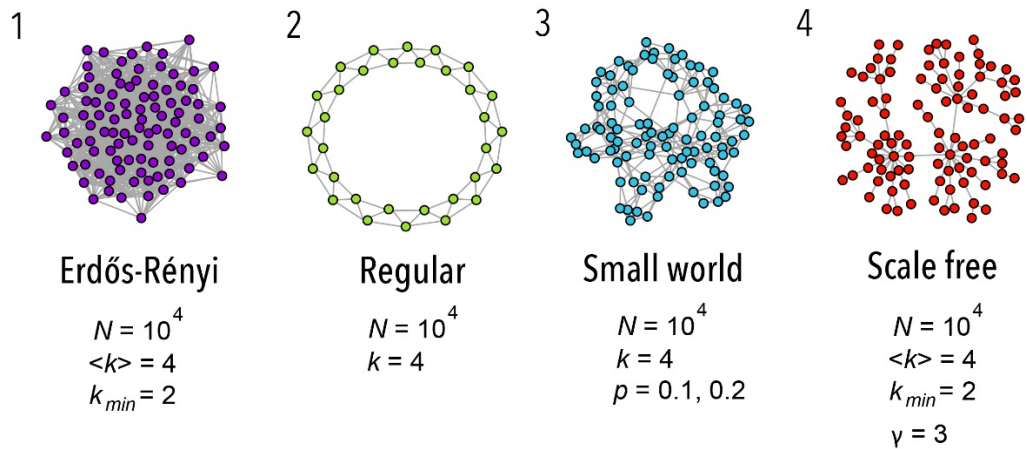

Note: We use four topologies to determine how actors influence each other via the ε signal. All networks were generated using the configuration model (Newman 2010), with the exception of the small world network, for which we used the Watts-Strogatz model (Watts and Strogatz 1998). The small world network rewires 1% of the ties of the regular network (we also run some simulations with p = 0.2, with qualitatively similar results). All networks have the same size ($N=10^4$) and the same average degree (for the Erdős–Rényi and the scale free networks) and degree (for the regular and small-world networks).

Figure 5. Maximum Number of Coordinated Actors as a Function of Time across ω Values

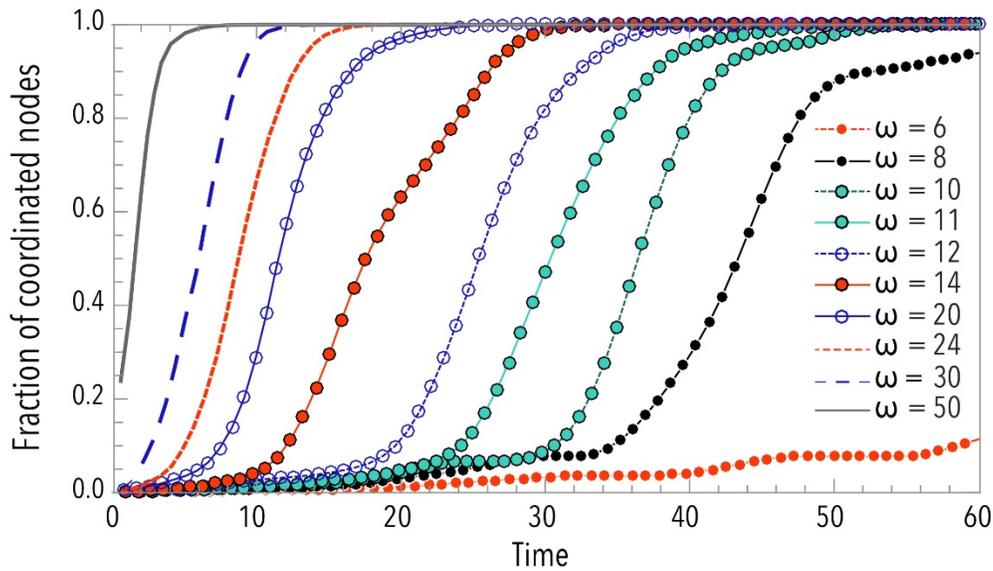

Note: The curves track the fraction of actors in the network that activate simultaneously. Simulations here run on a small world network with rewiring probability $p = 0.1$ and a fixed $\varepsilon = 0.01$. As expected, high ω values (which here is the same for all actors) lead to faster large-scale coordination. As ω decreases, the time to full coordination increases, down to values for which system-level coordination is unattainable (i.e. ω = 6, a condition under which only small clusters of coordinated nodes emerge).

Figure 6. The Impact of Social Influence on Large-Scale Coordination across ω Values

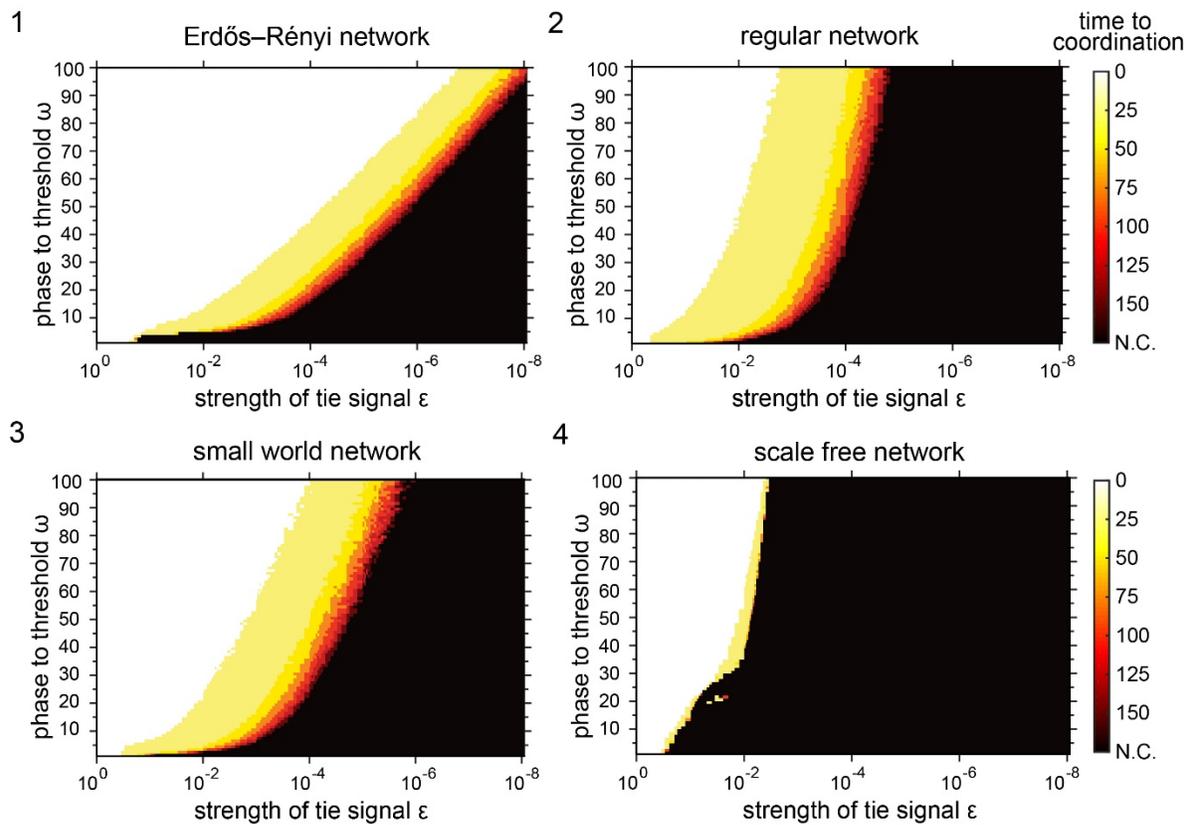

Note: The panels summarize coordination dynamics for different values of ω (the intrinsic motivation parameter) and ε (social influence strength) across the four network topologies. Every dot in the plots corresponds to a combination of parameters ε and ω; the distribution of ω and ε is homogenous across nodes and edges, respectively. The color scheme indicates how long it takes, for each combination, to reach large-scale coordination (which here we define as at least 75% of the nodes activating simultaneously); time is averaged over 100 realizations of the simulation. Lighter colors indicate earlier coordination, darker colors indicate later coordination; black signals that no large-scale coordination was possible. The findings suggest that random, homogenous networks are more conducive to large-scale coordination. Heterogeneous networks characterized by the presence of hubs (i.e. scale free networks) do not allow large-scale coordination soon after the social influence signal weakens. Small world networks are also more restrictive in the emergence of coordination than regular networks, in spite of the global shortcuts created by random rewiring (or because of them).

Figure 7. Heterogeneous Distribution of the Speed-to-Activation Parameter ω

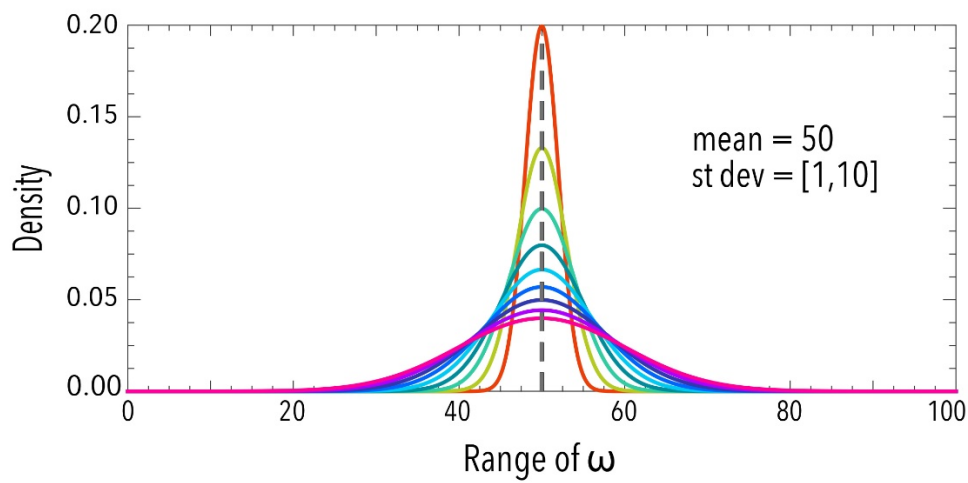

Note: In a second set of experiments we introduced actor heterogeneity by drawing the parameter ω from different normal distributions, centered around mean ω = 50 and with a standard deviation in the range σ = [1, 10].

Figure 8. The Impact of Actor Heterogeneity on Large-Scale Coordination across ε Values

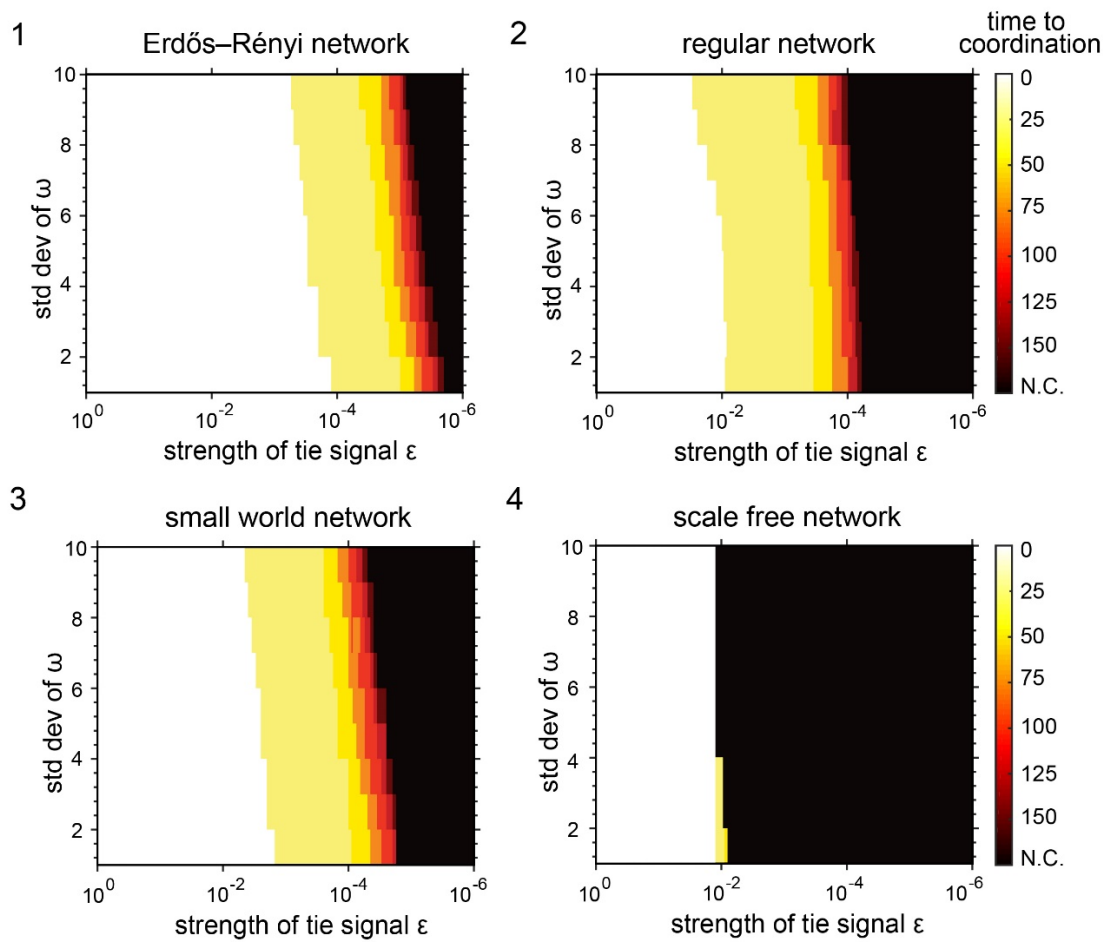

Note: The panels summarize coordination dynamics for different distributions of ω and ε values (social influence strength). The distribution of ω depends on the standard deviation (vertical axis); ε is homogenous across edges. The color scheme indicates, again, the time it takes to reach large-scale coordination (i.e. at least 75% of the nodes activating simultaneously); time is averaged over 100 realizations. The results show that, once more, all networks are less efficient in allowing large-scale coordination than the random benchmark provided by the Erdős–Rényi topology. Overall, low to mild heterogeneity increases the probability of global coordination, whereas high heterogeneity hinders it.